# How Specification Explicitness Shapes Failure in LLM-Generated Code

Elmira Onagh
York University
eonagh@my.yorku.ca

Maleknaz Nayebi
York University
mnayebi@my.yorku.ca

***Abstract*—When LLMs turn requirements into code, every underspecified constraint becomes behavior that the model must decide during generation. We test when specifications move beyond structure and begin to control generated behavior. In a controlled experiment, 72 participants implemented and revised Python tasks with an LLM under three levels of specification explicitness. The results show a clear break between lightweight structure and explicit behavioral obligations. Lightweight structure reduced hidden-test failures by only 8.7%. Requiring participants to specify constraints, invariants, negative requirements, edge cases, tie-breaking rules, and exception handling cut failures by 47.2% and reduced the number of distinct ways solutions failed by 20.5%. The largest drops were in failure modes tied to implicit obligations. Invariant violations fell by 66.1%, and unsupported assumptions fell by 59.5%. After a requirement change, specifications with explicit behavioral obligations produced regressions in more submissions, but each affected submission lost less previously correct behavior. Explicit specifications also changed how developers judged generated code. Without explicit obligations, participants overestimated generated-code correctness by about 23 percentage points. With explicit obligations, they produced better code but judged it more cautiously. Thus, the lesson is sharper than "better specifications help." Lightweight structure barely helped, while explicit behavioral obligations nearly halved failures, made regressions shallower, and improved calibration.**

## I. Introduction

Software failures rarely start in code alone. Decades of software engineering research show that defects introduced during requirements elicitation and specification drive downstream failures and maintenance costs [1], [2], [3], [4]. In traditional development, these defects often remain latent until implementation, testing, or deployment, where they surface as incorrect behavior, missed edge cases, or regressions. With LLM-assisted development, this latency collapses.

Large Language Models (LLMs) increasingly generate code from natural-language descriptions. In this setting, a prompt is no longer merely a set of instructions from a human developer; it becomes an operational requirements specification for an implementation agent. This collapses the distance from requirement to execution. It also removes intermediate points where humans traditionally clarify ambiguity, negotiate assumptions, and repair underspecification before implementation [5], [6], [7]. The risk of such development is assumption injection. When a specification is incomplete, the model may not ask what is missing; it may simply fill the gap. Recent evidence suggests that LLMs often proceed even when specifications omit constraints, priorities, or assumptions, producing behavior that is plausible but wrong [8], [9]. Ambiguity and underspecification can therefore become executable behavior immediately [10]. Classical requirements engineering argues that correctness depends on making assumptions, constraints, and invariants explicit [2], [5]. Yet this evidence largely assumes a human developer who interprets the specification before translating it into code. In LLM-assisted development, the specification is consumed directly by a model that generates executable artifacts. The resulting failures are not merely random generation errors [11], [12]; they can be specification-induced failures manifesting as brittle logic, missed edge cases, and regressions after small requirement changes caused by what the specification leaves unsaid [13].

Despite growing interest in LLM-assisted software development, we still lack empirical evidence on how specification explicitness shapes generated code. *Do LLMs reduce requirements debt by inferring missing details correctly, or amplify it by turning ambiguity into immediate implementation choices?* In this study, we treat requirements specification explicitness as a first-class experimental variable in LLM-assisted development. We conduct a controlled experiment with 72 participants to answer the following research questions:

**RQ1: How does specification explicitness shape failure structure in LLM-assisted development?**
We analyze the *types of failures* produced under different levels of specification explicitness. Using hidden tests and a predefined failure taxonomy, we evaluate whether explicitness changes frequency and types of program failure.

**RQ2: Which requirement elements most strongly predict correctness in LLM-assisted development?**
We examine which *requirement elements*, including constraints, invariants, edge cases, and acceptance criteria, are most strongly associated with hidden-test success. This identifies which parts of a specification matter most when requirements are consumed by a code-generating model.

**RQ3: How does specification explicitness affect behavioral regression and change amplification under requirement change?**
We study *robustness under change* by introducing a controlled requirement update and measuring pass-to-fail transitions (hidden tests that passed before the update but failed after it). This shows whether explicitness helps preserve previously correct behavior or whether changed requirements cause the LLM to break behavior that was already working.

**RQ4: How does specification explicitness affect developer calibration in LLM-assisted development?** In LLM-assisted development, failure is not only behavioral; it is also epistemic. A generated solution may look correct while silently violating constraints, invariants, or edge cases. We therefore ask whether specification explicitness helps developers judge correctness, not only produce correctness. We measure *developer calibration* as the gap between predicted and actual performance on the hidden test. This tests whether explicit specifications help developers evaluate generated code more accurately by making the expected behavior visible.

This paper makes the following contributions: (i) presents a controlled study of specification explicitness in LLM-assisted development, measuring effects on correctness, failure structure, regressions, and developer calibration, (ii) identifies which requirement elements are most associated with code correctness and which failure modes they suppress in LLM-generated code, and (iii) separates specification effects from prompting-effort effects using covariate models that control for prompt behavior, Python skill, and LLM familiarity.

## II. Related Work

We examined the literature relevant to our controlled study.

*Requirements Specification Quality.* Requirements Engineering (RE) is central to software quality because defects introduced early are costly to fix and often propagate into implementation and maintenance [1]. Zave frames RE as reasoning about domain assumptions, requirements, and specifications, making explicit assumptions a condition for correctness [2]. Ambiguity and omission in natural-language requirements are therefore recurring sources of defects [5], [14]. Prior work has addressed these risks through formal and model-based approaches, examining abstraction level, tool support, industrial applicability, and practical limitations [15], [16], [17], [18]. These principles are reflected in requirements standards such as IEEE 830 and ISO/IEC/IEEE 29148, which emphasize explicit constraints, negative requirements, invariants, and verifiable acceptance criteria [19], [20]. Specification quality also shapes evolution. Clear specifications support traceability, change impact analysis, regression control, and project outcomes [21], [22], [23], [24], [25]. Explicit constraints, invariants, and acceptance criteria can localize changes and reduce unintended side effects during maintenance [26], [27], [28]. Empirical studies further link well-defined Software Requirement Specifications (SRS) to fewer defects and failures [29], [30], and show that relationships among specification elements influence requirements quality, project success, team outcomes, and modeling productivity [31], [32], [33]. Our work extends this line of research to LLM-assisted development, where requirements are not only read by developers but also consumed by models and translated directly into code. This work builds on a broad body of prior research spanning requirements engineering, software analytics, release planning, software evolution, developer collaboration, and AI-assisted software engineering. We situate our contribution within this wider research context, drawing on its methodological perspectives, empirical investigations, and supporting research artifacts [34], [35], [36], [37], [38], [39], [40], [41], [42], [43], [44], [45], [46], [47], [48], [49], [50], [51], [52], [53], [54], [55], [56], [57], [58], [59], [60], [61], [62], [63], [64], [65], [66], [67], [68], [69], [70], [71], [72], [73], [74], [75], [76], [77], [78], [79], [80], [81], [82], [83], [84], [85], [86], [87], [88], [89], [90], [91], [92], [93], [94], [95], [96], [97], [98], [99], [100], [101], [102], [103], [104].

*LLMs in Software Engineering.* LLMs are increasingly used as coding assistants, autonomous agents, and interfaces between natural language and executable artifacts. Because most interactions are prompt-driven, recent work has studied prompt engineering for software development tasks [105], [106], [107], [108], [109], [110], [111]. A growing body of work examines how requirements appear in prompts and affect generated code quality [112]. Related studies show that LLM-generated code can diverge from natural-language intent [113], that formal RE and LLMs can inform each other [8], and that prompt under-specification causes fragile behavior that is difficult to repair through optimization alone [114]. Prompting methods such as AceCoder further show that explicit requirement understanding can improve code-generation performance

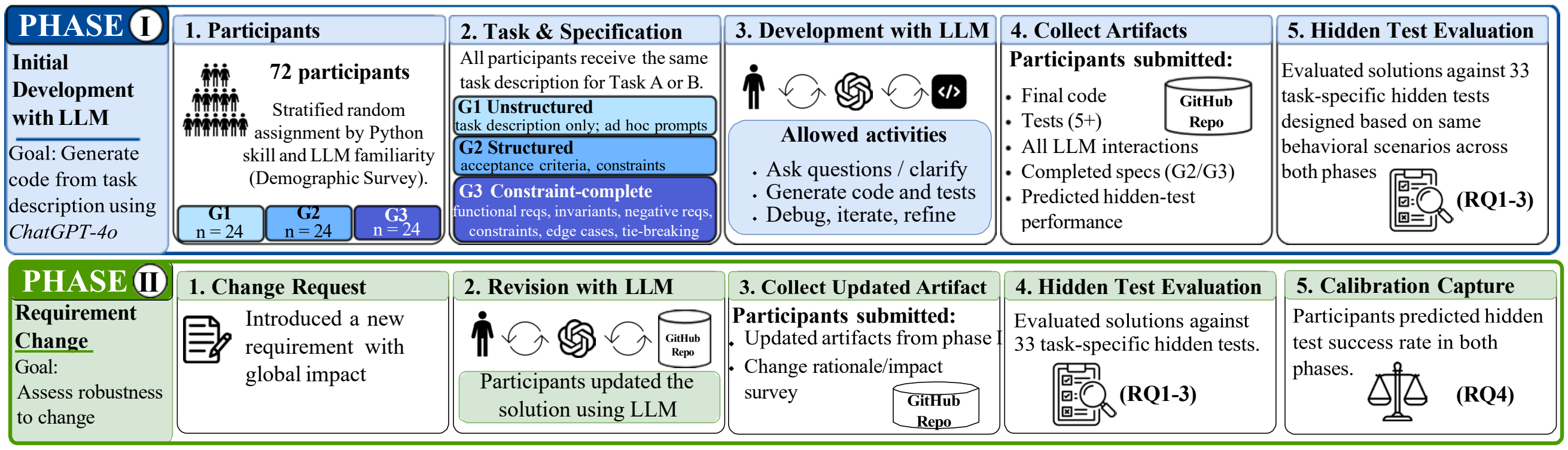

Fig. 1: Overview of the two-phase experimental protocol: initial LLM-assisted development (Phase I), followed by revision after a requirement change (Phase II).

**Task A: Meeting Slot Suggestion**

You are asked to implement a Python function that suggests possible meeting time slots for a given day. The function takes as input a list of existing calendar events for that day and a desired meeting duration. Its goal is to return a list of valid start times at which the meeting could be scheduled. A valid time slot should respect typical working hours and should not conflict with existing events. If no suitable time slots are available, the function should return an empty list. You may assume that time values are provided in a standard string format (e.g., "09:30"), and you are free to choose an appropriate internal representation.

**Task B: Resource Allocation Checker**

You are asked to implement a Python function that determines whether a set of resource requests can be satisfied given limited capacity. The function takes as input a list of requests and available resources and returns whether a valid allocation exists. The goal is to determine the feasibility of allocation based on the provided information. You may assume inputs are provided in a reasonable and consistent format.

Fig. 2: Task descriptions used in the experiment. These descriptions were identical across specification conditions.

[115]. Empirical studies also suggest that LLM-assisted failures often arise from missing assumptions and underspecified requirements rather than only syntactic or algorithmic errors [116], [117], [118]. Meanwhile, "vibe coding" describes exploratory, prompt-driven development with limited upfront specification [9], [119]. In contrast to this mostly descriptive and perception-based work, our study provides controlled evidence linking specification explicitness to failure modes, regressions, and calibration in LLM-assisted development.

## III. Experimental Design

To address the research questions, we conducted a controlled experiment with 72 developers. We employed a blocked between-subjects experimental design with two orthogonal factors: (i) *specification explicitness* as the primary independent variable, and (ii) *task* as a blocking factor to control for task-specific effects. The overview of our experimental procedure is depicted in Fig. 1.

### *A. Task Descriptions and Requirement Change*

For the experiment, we designed two `Python` tasks that were solvable within one hour but contained latent requirement ambiguities that had to be specified to avoid failures (Fig. 2). The tasks were comparable in scope, effort, and programming constructs, but targeted different requirement classes. Task A emphasized operational constraints, edge cases, and tie-breaking rules, whereas Task B emphasized global invariants, negative requirements, and consistency constraints.

This design supports RQ2 by identifying requirement classes associated with fewer failures while controlling for task difficulty. After the initial submission, participants received an additional task-specific requirement (Fig. 3). Each change was short in description but global in impact, requiring reasoning on constraints or invariants rather than a localized edit. Pilot runs confirmed the changes invalidated initial solutions while remaining feasible within the allotted time. All participants received the *same task description* for their assigned task. The experimental manipulation was the required specification level before LLM prompting: none in G1, structured in G2, and constraint-complete in G3 (more details in Section III-B).

### *B. Specifications*

We operationalized *specification explicitness* as the extent to which participants were required to externalize requirements before coding. The experiment used three levels: *unstructured* (G1), *Structured* (G2), and *Constraint-complete* (G3). Participants completed exactly one task under one specification level and were evenly distributed across the two tasks within each level. This blocked between-subjects design avoids carryover effects while enabling replication across tasks. Task descriptions were identical across groups; only the required level of specification explicitness varied.

**Unstructured Specification (G1).** Participants in G1 were not required to write an explicit specification before coding. They could interact with the LLM using ad hoc prompts derived from the task description. We used G1 as the baseline.

**Structured Specification (G2).** Participants in G2 received the same task descriptions as G1, but were required to complete a structured requirements template before coding (Fig. 4). The template asked participants to specify: (i) the goal, (ii) inputs and outputs, (iii) acceptance criteria in Given–When–Then form, and (iv) operational constraints and assumptions. Participants were instructed to use this specification as the primary reference when prompting the LLM.

**Constraint-complete Specification (G3).** Participants in G3 received the same task descriptions as G1 and G2, but were required to complete a more comprehensive specification before coding (Fig. 4). The template required participants to specify: (v) global invariants that must hold across all valid executions, (vi) negative requirements describing prohibited

**Task A: Phase II Requirement Change**

**New Requirement.** Meetings scheduled on Fridays must not start after 15:00 P.M. This requirement applies in addition to all previous behaviors. If a meeting would otherwise be valid but starts after 15:00 P.M. on a Friday, it must be excluded from the returned results. Update your solution to incorporate this requirement. If applicable, update your requirements specification before modifying code or tests.

**Task B: Phase II Requirement Change**

**New Requirement.** At least one resource must remain unallocated after assignment. This requirement applies in addition to all previous behaviors. An allocation that satisfies all other conditions but consumes all available resources is no longer valid. Update your solution to incorporate this requirement. If applicable, update your requirements specification before modifying code or tests.

Fig. 3: The new requirement introduced in Phase II for Tasks A and B to test for robustness of code generation.

**G2: Structured Specification Template**

**Goal.** Describe the primary objective of the system.
**Inputs.** List all inputs, their formats, and assumptions.
**Outputs.** Describe the output format and ordering.
**Acceptance Criteria**: Given ... When ... Then ...
**Constraints.** List constraints that restrict valid behavior.
**Assumptions.** List any assumptions about the system.

**G3: Constraint-complete Specification Template**

**Goal.** Describe the primary objective of the system.
**Inputs.** List all inputs, their formats, and assumptions.
**Outputs.** Describe the output format and ordering.
**Functional Reqs.** List the system's functional requirements.
**Constraints.** List constraints that restrict valid behavior.
**Invariants.** List properties that must hold for all valid executions.
**Negative Reqs.** List behaviors the system must not exhibit.
**Edge Cases.** List boundary conditions and exceptional scenarios.
**Tie-Breaking Rules.** List explicit priority among valid outcomes.

Fig. 4: All participants received the same task description; G2 and G3 completed structured templates before implementation.

behavior, (vii) edge cases and boundary conditions, and (viii) tie-breaking or prioritization rules among valid outcomes.

For G2 and G3, participants were not allowed to introduce new behavior without first updating the specification. This design isolates the effect of specification explicitness on failure modes, robustness to change, and developer confidence in LLM-assisted development.

### C. Participants

We recruited participants through convenience sampling from a cohort of 300 individuals who had completed requirements engineering coursework within the last five years. The coursework covered requirement types, processes, constraints, invariants, and acceptance criteria. Of the 76 respondents, 72 completed the study. Participants completed a demographic survey at the beginning of the experiment, reporting their `Python` proficiency and familiarity with LLM-based development tools (Fig. 5). Using stratified randomization [120] on these measures, we assigned them to six balanced groups: three specification levels (G1–G3) crossed with two tasks (A and B), with 12 participants per group. There was no compensation for the experiment.

### D. Experiment Process

The controlled experiment was conducted in a supervised session and consisted of two phases.

**Phase I (Initial Development).** Participants received Task A or B (Fig. 2) and completed a demographic survey (Fig. 5). The task descriptions were identical across conditions; only the required pre-implementation specification differed. G1 proceeded directly to implementation, whereas G2 and G3 first produced a specification artifact (Fig. 4). All participants implemented the task, wrote at least five tests, and used the free version of `ChatGPT-4o` for reasoning, coding, debugging, and test generation. Before they submitted a GitHub repository containing specifications, code, tests, and prompt transcripts, they confirmed that their code passed their own tests and reported their expected hidden-test performance.

**Phase II (Requirement Change).** After the initial submission, participants received a new task-specific requirement (Fig. 3) and completed an impact-prediction survey on the expected scope of change. G2 and G3 participants then updated their specifications before revising the code. All participants updated their code and tests, committed the changes, and logged any further `ChatGPT-4o` interactions. Finally, they completed a change-rationale survey describing their modifications and confidence in passing the hidden tests. Participants received no feedback before or during the experiment.

## IV. Data Collection & Pre-Processing

We collected participants' GitHub repositories containing the artifacts produced across both phases: source code, tests, G2/G3 specifications, and LLM prompt transcripts. We also collected survey data, including demographics, Phase I performance predictions, Phase II impact predictions, and change rationales. We did not impose commit-frequency or version-control requirements beyond the predefined checkpoints. We processed survey responses using extraction schemas implemented with AIRPARSER [121], followed by manual checks of the extracted fields. All anonymized data and extended analysis, including extraction schemas and survey templates, are available in our replication package [122].

## V. Measures

We measure correctness, types of failures, robustness, requirement impact, and calibration using the collected data.

**Hidden-test outcomes.** We measure behavioral correctness by executing each submission against a hidden test suite, following standard code-generation evaluation practice [13], [123], [124]. We designed 33 hidden tests per task before evaluation, based on stated functionality in description (Fig. 2) and plausible domain rules. We reused the same scenarios across phases. Phase II changed only the oracle for affected scenarios; all others retained their Phase I oracle. For participant $i$, let $T_i$ be the hidden-test set and $h_{it} \in \{0, 1\}$ indicate whether

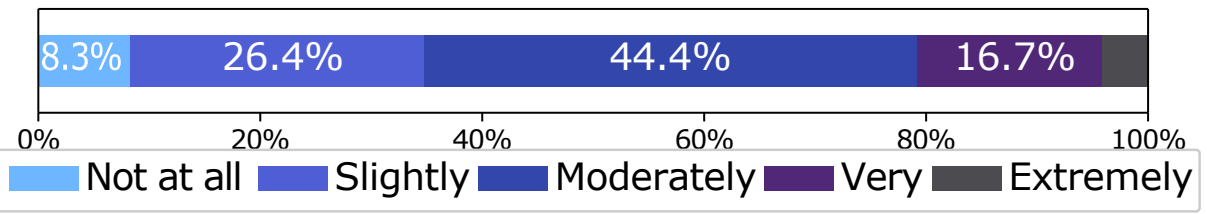

**(a) Participants Self-reported Python Familiarity**

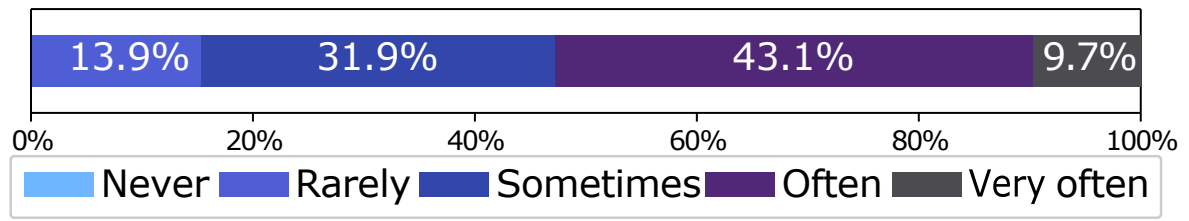

**(b) Participants Self-reported LLM Use**

Fig. 5: Python proficiency and LLM use of 72 participants.

test $t$ failed (1 = failed, 0 = passed). We define failed tests $F_i$, passed tests $P_i$, and full correctness $FC_i$:

$$F_i = \sum_{t \in T_i} h_{it}, \quad P_i = |T_i| - F_i, \quad FC_i = \begin{cases} 1 & \text{if } F_i = 0 \\ 0 & \text{otherwise} \end{cases}$$

These measures support RQ1 and RQ2.

**Failure modes.** To capture how submissions fail, each hidden test is labeled with one requirement-related class: *constraint omission (CO)*, *invariant violation (IV)*, *tie-breaking failure (TB)*, *edge-case failure (EC)*, or *unsupported assumption (UA)*. This follows defect and failure-classification practice, where failures are grouped by semantic cause [125], [126]. Let $C = \{CO, IV, TB, EC, UA\}$ and let $T_{ic} \subseteq T_i$ denote the hidden tests targeting category $c$. For participant $i$ and category $c$, we compute:

$$f_{ic} = \sum_{t \in T_{ic}} h_{it}, \quad I_{ic} = 1[f_{ic} > 0], \quad NF_{ic} = \frac{f_{ic}}{|T_{ic}|}, \quad D_i = \sum_{c \in C} I_{ic}.$$

Here, $f_{ic}$ is the category-specific failure count, $I_{ic}$ is category occurrence, $NF_{ic}$ is failure frequency normalized for unequal numbers of failures across categories, and $D_i$ is failure-mode diversity capturing the breadth of distinct failure mechanisms exposed by a solution. These measures support RQ1.

**Requirement effects.** To identify which specification elements are associated with correctness, we coded whether each participant's specification included *functional requirements*, *invariants*, *negative requirements*, *constraints*, *acceptance criteria*, *exception handling*, and *tie-breaking rules*. These categories follow standard guidance on behavior, constraints, assumptions, verifiability, and acceptance criteria [19], [20]. We then relate each element $c$ to each failure mode $f$ using a risk-difference contrast [127]:

$$\Delta_{c,f} = P(Fail_f \mid Req_c = 0) - P(Fail_f \mid Req_c = 1),$$

where $Fail_f$ indicates the presence of failure mode $f$. Positive values indicate a lower likelihood of failure when requirement element $c$ is present. This descriptive contrast supports RQ2.

**Robustness under change.** To evaluate robustness after the Phase II requirement change (RQ3), we track regressions as pass-to-fail transitions: hidden tests that passed after the initial implementation but failed after the change. This follows regression-testing practice, where regressions capture previously working behavior that breaks after a program modification [128], [129]. For each participant, we report *regression occurrence* (whether at least one regression occurred), *regression severity* (count of regressions), and *regression rate* (the proportion of previously passing tests that regressed).

**Developer calibration.** We measure developer calibration as the signed gap between predicted and observed hidden-test performance. After each phase, participants predicted the performance of their solution using a five-point confidence scale. We map each response to the midpoint of its percentage range and compare it with the observed hidden-test pass rate. Positive values indicate overconfidence, where participants overestimate correctness; negative values indicate underconfidence, where participants underestimate it. This measure supports RQ4 and follows calibration research comparing subjective confidence with observed accuracy [130], [131].

## VI. Analysis

We treat specification explicitness as the main experimental factor and task as the secondary one. Across RQs, models report effect sizes and uncertainty, while descriptive analyses explain how specification shapes outcomes. Pairwise comparisons are adjusted within each RQ.

### A. RQ1: Correctness and failure structure.

We compare *failure count*, *passed tests count*, and *full correctness* across conditions. Counts are analyzed using one-way ANOVA with Tukey HSD comparisons [132]; full correctness is reported descriptively because sparse successes cause logistic model separation. We also compare normalized *failure frequencies* for each failure class and per-submission *failure-mode diversity* using ANOVA. We report both statistical significance and practical importance: partial eta squared ($\eta_p^2$) for overall group effects, Hedges' $g$ for pairwise contrasts, and 95% confidence intervals for both [133], [134]. Because failure counts are over-dispersed, we also fit a negative binomial regression [135] for the number of failed hidden tests We report Incidence Rate Ratios (IRRs), with values below 1 indicating fewer expected failures than in the unstructured condition, Holm-adjusted $p$-values, and 95% confidence intervals.

### B. RQ2: Requirement elements and failure reduction.

We model hidden-test success using binomial logistic regression, with passed versus failed hidden tests as the outcome and requirement-element indicators as predictors. We report odds ratios, 95% confidence intervals, and adjusted $p$-values. To connect these effects to concrete failure mechanisms, we use the requirement–failure contrast measure (Section V) and visualize it as a heatmap. We interpret RQ2 as associative rather than causal: requirement elements may co-occur with other participant behaviors, and their presence does not guarantee precision or use during prompting.

### C. RQ3: Robustness under requirement change.

We examine whether specification explicitness affects robustness following the Phase II requirement update. Given the rarity of regressions, we treat RQ3 as a change-profile analysis. We report group-level and task-stratified descriptive statistics and model regression occurrence using logistic regression with specification condition and task as primary predictors. We report odds ratios with exact or profile-likelihood confidence intervals and interpret results conservatively, emphasizing effect direction, practical magnitude, and severity profiles rather than statistical significance [136], [137], [138].

### D. RQ4: Developer confidence calibration.

We compute a calibration gap for each participant (Section V). We summarize these gaps by specification condition and phase, then fit a mixed-effects model to test whether calibration differs across conditions and changes after the Phase II requirement update. The model controls for task and

TABLE I: RQ1 results by specification condition, showing hidden-test correctness, adjusted failure-count effects, normalized failure modes, and failure-mode diversity.

| Spec. | Mean fail. /33 | Full corr. | Adj. IRR [95% CI] | $p$ / $p_{\text{Holm}}$ | CO | IV | TB | EC | UA | Mean fail. div. |
|---|---|---|---|---|---|---|---|---|---|---|
| G1 Unstructured | 21.54 | 0.00% | 1.00 [–] | – | 76.49 | 92.71 | 8.33 | 48.15 | 58.33 | 4.04 |
| G2 Structured | 19.67 | 0.00% | 0.95 [0.50, 1.82] | 0.884 / 1.000 | 71.06 | 85.42 | **0.00** | 47.92 | 49.54 | 4.00 |
| G3 Constraint-complete | **11.38** | **4.17%** | 0.48 [0.24, 0.93] | 0.029 / 0.205 | **51.36** | **31.46** | **0.00** | **31.02** | **23.61** | **3.21** |

*Notes:* fail. div. = failure-mode diversity. Failure-mode values are normalized percentages. Specification significantly affected failure count, passed tests, and failure-mode diversity ($p < 0.001$). The adjusted model found fewer expected failures for G3 than G1, although the effect was not significant after Holm correction ($p_{\text{Holm}} = 0.205$).

includes participants as a random effect to account for repeated judgments. We report estimated contrasts, 95% confidence intervals, and Benjamini–Hochberg (BH)-adjusted $p$-values.

## VII. Results

We present results from the controlled experiment by RQs.

### A. Specification explicitness impact on failure structure (RQ1)

*Constraint-complete specifications reduce failure frequency.* Table I shows that specification explicitness matters when it exposes the behavioral obligations the LLM must preserve. Mean hidden-test failures decreased from 21.54 out of 33 in the unstructured condition (G1) to 11.38 in the constraint-complete condition (G3), where participants specified a detailed specification (Section III-B). The structured condition (G2), which required only constraints and assumptions, produced a smaller reduction to 19.67 failures (Fig. 6-a). Thus, *the gain is not from structure alone. It appears when the specification makes hidden behavioral rules explicit enough that the LLM no longer has to infer them.* Full correctness remained rare: one G3 submissions (4.17%) passed all 33 tests, compared with none in G1 or G2. The group comparison confirms that specification explicitness changes both failures and successes. Failure counts differ across three groups, $F(2, 69) = 26.77$, $p < 0.001$, $\eta^2_{\varepsilon} = 0.44$, and the same effect appears for passed tests. The effect is large, with 44% of the observed variation in hidden-test failures attributable to differences among the three specification conditions. We fit an adjusted negative binomial model for hidden-test failures (Section VI), as failures are non-negative counts. The model controlled for *task*. The adjusted results followed the same direction: G3 was associated with fewer expected failures than G1, IRR = 0.48, 95% CI $[0.24, 0.93]$, $p = 0.029$, whereas G2 did not differ from G1, IRR = 0.95, 95% CI $[0.50, 1.82]$, $p = 0.884$. Overall, the unadjusted and adjusted analyses show the same pattern: lightweight structure is insufficient, while constraint-complete specifications reduce failures.

*The gains are concentrated in failures caused by missing behavioral obligations.* Table I shows that G3 reduces failures that occur when the LLM must infer unstated behavior: constraint omissions fall from 76.49% to 51.36%, invariant violations from 92.71% to 31.46%, unsupported assumptions from 58.33% to 23.61%, and edge-case failures from 48.15% to 31.02%. Tie-breaking failures disappear in both G2 and G3, indicating that even a lightweight structure can resolve local ordering or selection rules. Broader reductions appear only in G3, where more behaviors are explicit. Constraint-complete specifications (G3) also reduce failure-mode diversity. Submissions in G1 and G2 fail across a similar number of classes on average (4.04 and 4.00), compared with 3.21 in G3 (Fig. 6-b). The difference is significant, $F(2, 69) = 10.46$, $p < 0.001$, $\eta^2_{\varepsilon} = 0.23$. Tukey HSD shows lower diversity in G3 than in both G1 and G2 ($p < 0.001$), while G1 and G2 do not differ ($p = 0.978$). Thus, *constraint-complete specifications reduce not only the count of failures but also the number of distinct ways LLM-assisted solutions fail.*

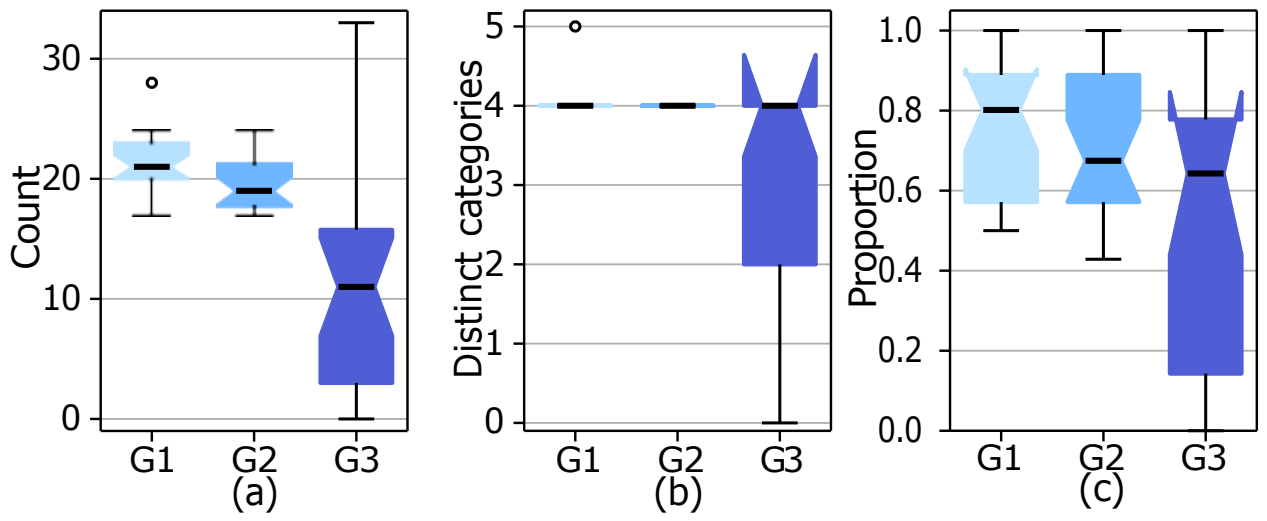

Fig. 6: Distribution of (a) failure counts, (b) # of distinct failure categories, and (c) proportion of constraint-omission hidden tests failed across groups (RQ1).

Explicit behavioral obligations shape LLM failure. LLM failures are not random hallucinations; they followed what the specification left implicit. Lightweight structure (G2) resolved simple ambiguities but was close to the unstructured condition (G1). Constraint-complete specifications (G3) reduced mean failures by about half, lowered failure-mode diversity, and reduced failures involving missing constraints, invariants, edge cases, and unsupported assumptions. The lesson is not "write more," but "make the right obligations explicit." Specifications determine which behaviors are fixed and which remain for the LLM to infer.

### B. Requirement elements that predict correctness (RQ2)

*Not all requirement elements predict correctness equally.* Table II reports binomial logistic models relating each requirement element to hidden-test success. *Functional requirements* show the strongest positive association (OR = 2.80, $p_{\text{BH}} < 0.001$), followed by *invariants* and *negative requirements* (both OR = 2.62, $p_{\text{BH}} < 0.001$). *Tie-breaking rules* (OR = 1.99, $p_{BH} < 0.001$) and *exception handling* (OR = 1.48, $p_{BH} < 0.001$) also show smaller but positive associations. *Constraints* are positive but not significant (OR = 1.36, $p_{BH} = 0.384$), likely because they appear in most G2/G3 specifications and provide limited contrast. *Acceptance criteria* are negatively associated with success (OR = 0.37, $p_{BH} < 0.001$). We do not interpret this as evidence that they reduce correctness; they may appear more often in harder specifications with more expected behaviors. *Correctness is therefore associated not simply with having a specification, but with which requirement elements the specification makes explicit.*

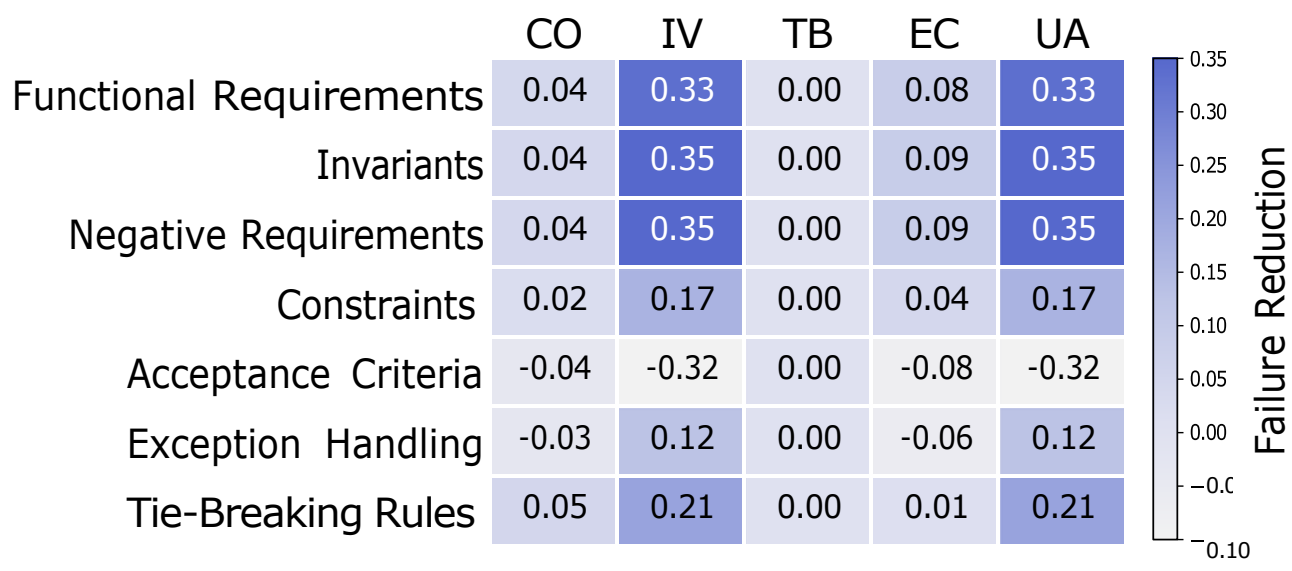

Fig. 7: Requirement elements and failure-mode reduction. Cells report risk differences; positive values indicate lower failure likelihood when the element is present (RQ2).

*Requirement elements reduce different failure modes.* Fig. 7 shows that elements defining global behavior have the strongest reductions. *Invariants* and *negative requirements* show the largest reductions in invariant violations and unsupported assumptions (risk difference = 0.35), followed by *functional requirements* (0.33). Thus, elements specifying what must hold, what must not occur, and what behavior must be implemented are most effective at suppressing LLM failures. Localized elements show narrower effects. *Tie-breaking rules* reduce invariant violations and unsupported assumptions by 0.21, while *exception handling* shows smaller reductions of 0.12. *Acceptance criteria* have negative risk differences across several modes, consistent with the logistic results. We interpret this as a complexity signal because acceptance criteria appear more often in harder specifications, not as evidence that they directly reduce correctness. Overall, *elements that constrain global behavior show the strongest predictive signal for hidden-test success and failure-mode reduction.*

TABLE II: Predictive power of requirement elements for hidden-test success (RQ2).

| **Requirement Element** | **Odds Ratio** | **95% CI** | $p_{BH}$ |
|---|---|---|---|
| Functional Requirements | 2.80× | [2.29, 3.44] | < 0.001 |
| Invariants | 2.62× | [2.14, 3.21] | < 0.001 |
| Negative Requirements | 2.62× | [2.14, 3.21] | < 0.001 |
| Tie-Breaking Rules | 1.99× | [1.62, 2.44] | < 0.001 |
| Exception Handling | 1.48× | [1.20, 1.83] | < 0.001 |
| Constraints | 1.36× | [0.68, 2.72] | 0.384 |
| Acceptance Criteria | 0.37× | [0.30, 0.45] | < 0.001 |

*Notes:* p-values adjusted using Benjamini–Hochberg correction.

> Not all requirement elements carry equal value in LLM-assisted development. Correctness depends less on the presence of a specification than on what it makes explicit. Invariants, negative requirements, and functional requirements provide the strongest signals for hidden-test success and broad failure reduction. The most useful specification is the one that makes failure-relevant behavior explicit and constrains global behavior.

### C. Specification explicitness and change robustness (RQ3)

*Specification explicitness does not monotonically reduce regression under change; it reshapes the regression profile.* Table III summarizes pass-to-fail transitions after the requirement change. Regressions are uncommon: only eight of 72 submissions exhibit at least one previously passing test that later fails (Section V). We therefore treat RQ3 as a change-profile analysis instead of a significance-driven comparison. Regression occurrence is identical in G1/G2 (2/24; 8.3%) and higher in G3 (4/24; 16.7%). The task-adjusted model shows the same pattern: G2 is similar to G1 (OR = 1.00), while G3 has a higher but imprecise estimate (OR = 2.22); neither contrast is statistically significant. Given the small number of events, these estimates describe the pattern of change behavior rather than precisely estimating regression risk.

*More regressions do not mean more damage.* Severity reveals a trade-off. G3 has the highest regression occurrence, with 4/24 affected submissions. But the regressions are shallower. G2 affects only 2/24 submissions, yet loses the most previously passing behavior, with an average of 0.38 pass-to-fail transitions and a normalized regression rate of 2.38%. G3 shows fewer behavioral losses, with 0.17 pass-to-fail transitions and a 0.70% regression rate. *Specification explicitness therefore does not simply make changed code safer. It changes what breaks.* Constraint-complete specifications (G3) expose more behavior to the change, but limit the behavioral loss when regression occurs. Structured specifications affect fewer submissions, but when they fail, they fail deeper.

> Explicit specifications are double-edged under change. They do not simply make revisions safer; they change what breaks. Constraint-complete specifications (G3) affect more submissions but lose less previously passing behavior, whereas structured specifications (G2) affect fewer submissions but lose more behavior when regressions occur. This profile suggests that specification explicitness shapes which behaviors the LLM preserves under change: earlier obligations may remain influential as requirements evolve.

### D. Specification explicitness and confidence calibration (RQ4)

*Explicit specifications make developers less likely to overtrust generated code.* Table IV summarizes developer calibration across tasks and phases. Calibration matters because LLM-generated code can appear plausible yet still miss hidden constraints, invariants, or edge cases. This risk is most evident in the unstructured condition (G1), where participants persistently overestimated correctness by 22.78 percentage points after initial development and by 22.73 percentage points after the requirement change. Structured specifications (G2) reduced this gap to 5.65 points after initial development and shifted participants toward underconfidence after the change, $-17.43$ points. Constraint-complete specifications (G3) made participants conservative in both phases, with calibration gaps of $-22.20$ and $-18.22$ points. Thus, explicit specifications did not make developers more confident. They made developers less likely to overtrust plausible generated code by making missing obligations easier to see.

*The mixed-effects model shows that explicit behavioral obligations make developers less overconfident in LLM-generated code.* During initial development, the mixed-effects model shows that G3 differs sharply from G1, reducing overconfidence by 45.33 percentage points, 95% CI [-60.87, -29.80], $p_{\mathrm{BH}} < 0.001$. The G2–G1 contrast is smaller and uncertain, -13.20 points, 95% CI [-28.73, 2.34], $p_{\mathrm{BH}} = 0.154$. After the requirement change, G2 shows an additional reduction in overconfidence, -24.75 points, 95% CI [-41.54, -7.96], $p_{\mathrm{BH}} = 0.015$, while G3 remains stable, 0.01 points, $p_{\mathrm{BH}} = 0.999$. These results suggest that *explicit specifications expose more reasons to doubt generated code, reducing overestimation even when the generated solution appears plausible.* Overall, explicit specifications reduce overestimation but shift developers toward more conservative judgments.

> Developers overtrust LLM-generated code when requirements leave constraints and invariants implicit. When participants did not extract specifications before prompting (G1), they overestimated hidden-test performance by about 23% in both phases. When they exhausted constraints, invariants, negative requirements, and edge cases (G3), they were more conservative in judging generated code as correct, even though the code performed better. Explicit specifications make hidden obligations checkable, guiding generation and giving developers a basis for trusting the code.

## VIII. Discussion

Our findings show that specification explicitness is not merely a documentation practice in LLM-assisted development. It shapes the behavioral obligations available to the model, the assumptions embedded in generated code, the way systems respond to change, and how accurately developers judge correctness. Across the RQs, the central lesson is that failures often arise when requirements are absent from, or lost between, the specification, prompt, tests, and code.

TABLE III: Post-requirement-change regression profile (RQ3).

| Spec. | Reg. occur. | Occur. rate | OR [95% CI] | Sev. | Reg. rate |
|---|---|---|---|---|---|
| G1 Unstructured | 2/24 | 8.3% | 1.00 [–] | 0.08 | 0.58% |
| G2 Structured | 2/24 | 8.3% | 1.00 [0.11, 8.99] | **0.38** | **2.38%** |
| G3 Constraint-complete | **4/24** | **16.7%** | 2.22 [0.39, 17.35] | 0.17 | 0.70% |

*Note.* ORs are task-adjusted ($N$ = 72). Given eight regression events, we emphasize profiles and effect magnitude.

TABLE IV: Dev. calibration by specification condition (RQ4).

| | Phase I | | | Phase II | | |
|---|---|---|---|---|---|---|
| **Spec.** | **Pred.** | **Act.** | **OC** | **Pred.** | **Act.** | **OC** |
| G1 | 3.38 | 2.13 | 22.78 | 3.63 | 2.33 | 22.73 |
| G2 | 2.77 | 2.55 | 5.65 | 3.30 | 3.96 | -17.43 |
| G3 | 2.67 | 3.71 | -22.20 | 3.30 | 4.13 | -18.22 |

*Mixed model for overconfidence*

| Effect | Est. [95% CI] | $p_{\mathrm{BH}}$ |
|---|---|---|
| G2 vs. G1 (Phase I) | -13.20 [-28.73, 2.34] | 0.154 |
| G3 vs. G1 (Phase I) | -45.33 [-60.87, -29.80] | <0.001 |
| Phase II vs. I (G1) | 3.30 [-8.68, 15.29] | 0.673 |
| G2 × Phase II | -24.75 [-41.54, -7.96] | 0.015 |
| G3 × Phase II | 0.01 [-16.69, 16.72] | 0.999 |
| Task B | -13.34 [-23.75, -2.92] | 0.032 |

*Notes:* Pred./Act. are mean categories on a five-point scale. OC is the signed calibration gap in percentage points; positive values indicate overconfidence and negative ones underconfidence. Estimates are adjusted for tasks. $p_{\mathrm{BH}}$ gives Benjamini–Hochberg-adjusted $p$-values. Reference levels are G1, initial development (Phase I), and Task A.

### A. Developer experience did not explain specification effects

Participants reported `Python` proficiency and LLM familiarity on five-point scales (Fig. 5). We used both measures for stratified assignment (Section III-C) and included them as covariates in the adjusted analyses for RQ1, RQ3, and RQ4.

Including these covariates did not materially change the observed differences, as the effects remained in the same direction and did not meaningfully shrink across RQs. This suggests that the observed effects were driven by the study conditions, rather than by differences in programming proficiency or prior LLM experience. In RQ1, constraint-complete specifications remained associated with fewer expected failures, while structured specifications remained similar to the unstructured condition (Section VII-A). In RQ3, the expanded model produced similar estimates (G2 vs. G1: OR = 0.96; G3 vs. G1: OR = 2.38; Section VII-C). In RQ4, neither `Python` proficiency (estimate = 5.64, 95% CI [$-0.08$, 11.36], $p_{\mathrm{BH}}$ = 0.106) nor LLM familiarity (estimate = 2.56, 95% CI [$-3.34$, 8.47], $p_{\mathrm{BH}}$ = 0.527) was significantly associated with calibration. Thus, *measured experience did not explain the failure, regression, or calibration patterns, although self-reports may not capture all aspects of expertise.*

### B. Prompting effort also did not explain the specification effect

One possible explanation for the lower failure rates in G3 is that participants simply interacted more with the LLM. To test this, we measured LLM interaction effort using *prompt count*, *total and average prompt length*, and counts of *clarification*, *debugging*, *code-generation*, and *test-generation* prompts. Prompt types were generated with fine-tuned `openai/gpt-oss-120b` and were validated by two authors on a stratified 20% sample, with substantial agreement (Cohen's $\kappa$ = 0.81). We included these measures as covariates in RQ1, RQ3, and RQ4 to assess whether specification

TABLE V: Specification reuse within G2/G3. IRR values < 1 indicate fewer expected failures for the higher reuse.

| Prompt–spec. reuse | $n$ | Mean adh. | Mean fail. /33 | Fail. div./5 | Adj. IRR [95% CI] |
|---|---|---|---|---|---|
| Lower reuse | 13 | 0.091 | 19.46 | 3.85 | 1.00 [–] |
| Higher reuse | 35 | 0.671 | 14.06 | 3.51 | 0.76 [0.32, 1.83] |

*Note.* G2/G3 participants were split at the median adherence score (0.50). The adjusted model found fewer expected failures with higher reuse, but the effect was not significant (IRR = 0.76, 95% CI [0.32, 1.83], $p = 0.543$, $p_{\text{Holm}} = 1.000$).

explicitness remained associated with failures, regressions, and calibration after accounting for interaction effort [139], [140].

Across 718 LLM interactions, participants submitted 341 prompts in Phase I and 377 in Phase II. Mean prompt count increased from 4.87 to 5.24 per participant after the requirement change, while mean total prompt volume increased from 598.6 to 937.1 words per participant. Prompt purposes also shifted: code-generation prompts increased from 1.51 to 2.01 per participant, whereas debugging prompts decreased from 0.94 to 0.45. G3 participants used fewer debugging prompts than the other conditions. Detailed prompt-analysis results are available in the replication package [122]. Including prompt effort did not materially change the findings. In RQ1, G3 remained associated with fewer failures, while G2 remained close to G1 (Table I, adjusted $p_{\text{Holm}}$). RQ3 estimates were also similar after adjustment, with OR = 0.96 for G2 vs. G1 and OR = 2.38 for G3 vs. G1. In RQ4, neither prompt effort nor prompt type remained associated with overconfidence after correction. The specification effect was therefore not a more-prompting effect. Participants did not perform better merely because they wrote more prompts, longer prompts, or different kinds of prompts.

### C. *Specifications helped when obligations reached the prompt*

Prompting effort captures how much participants interacted with the LLM. Adherence captures something different: whether the written specification was actually used in that interaction. For G2 and G3, we measured whether specified obligations were carried into prompts. Two annotators coded 20% of transcripts for reuse or omission of *constraints*, *edge cases*, *invariants*, and *test obligations*. These labels were used to tune an `openai/gpt-oss-120b` classifier, which achieved 75.12% agreement with manual labels on a separate 10% audit. We defined *adherence score* as the proportion of specified elements referenced at least once, and *omission rate* as the proportion never referenced [141], [142].

Within G2/G3, we split participants at the median adherence score, 0.50, and fit models adjusted for task, prompting effort, and developer experience. We also related adherence to the RQ2 requirement-element results, explored its role in RQ3 regressions, and examined its association with RQ4 calibration. For RQ2, we checked whether the requirement elements associated with hidden-test success were also carried into prompts. For RQ3, we tested whether higher adherence was associated with fewer pass-to-fail transitions after the requirement change. For RQ4, we tested whether higher adherence was associated with a smaller predicted-minus-actual correctness gap.

Adherence was higher in G3 than in G2. Participants in G3 referenced 60.0% of specified elements on average, compared with 45.0% in G2. Higher adherence was directionally associated with better correctness. In RQ1, higher-adherence participants had fewer failures, 14.06 vs. 19.46, and lower failure-mode diversity, 3.51 vs. 3.85. The adjusted effect, however, was not significant, IRR = 0.76, 95% CI [0.32, 1.83], $p = 0.543$ (Table V).

Fig. 8 shows a positive but noisy association with pass rate, supporting the RQ2 interpretation that requirement elements matter most when they are operationalized rather than left only in the written artifact. Adherence did not explain the RQ3 frequency–severity trade-off. The clearest signal concerned developer judgment in RQ4. Greater adherence was associated with a 7.77-percentage-point lower calibration gap, 95% CI $[-14.65, -0.89]$, $p = 0.027$. Because these analyses were exploratory, we treat them as mechanism evidence rather than confirmed effects. *Together, the results suggest that explicit specifications help when their obligations are not only documented, but carried into the LLM interaction.*

### D. *Explicitness reduces ambiguity but must be maintained under change*

The results qualify the intuition that more explicit specifications always make change safer. In initial development, constraint-complete specifications clearly reduced ambiguity. G3 cut hidden-test failures from 21.54 in G1 and 19.67 in G2 to 11.38, a 47.2% reduction over G1, while lightweight structure reduced failures by only 8.7%. G3 also reduced the number of distinct failure modes by 20.5%. Explicit obligations therefore helped most before the requirement changed, when constraints, invariants, edge cases, and prohibited behaviors had to be fixed rather than inferred. After the requirement change, explicitness changed the regression profile rather than eliminating regression. Regressions were rare overall, but G3 had more affected submissions than G2, 16.7% vs. 8.3%. The affected G3 submissions, however, lost less previously correct behavior. G2 averaged 0.38 pass-to-fail transitions, while G3 averaged 0.17. Compared with G2, G3 reduced the average number of pass-to-fail transitions by 55.3% and the normalized regression rate by 70.6%. Thus, constraint-complete specifications did not make change regression-free. They made regressions shallower.

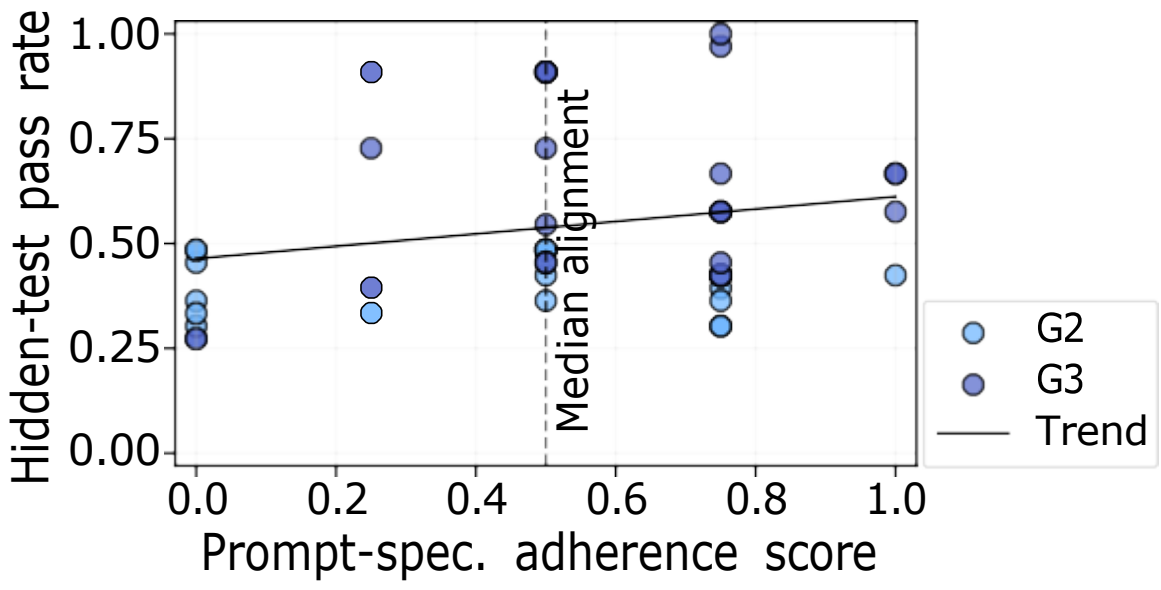

Fig. 8: Prompt–specification adherence and hidden-test pass rate for G2/G3. Higher adherence indicates greater reuse of specification content in LLM prompts.

*This pattern suggests an ambiguity–maintenance tradeoff. Under-specified prompts leave behavior for the model to infer, increasing the risk of incorrect initial implementations. Explicit specifications reduce this inference burden, but their obligations must also be updated consistently as requirements evolve.* Our adherence analysis provides exploratory support for this. G3 participants carried more specified elements into prompts than G2 participants, 60% vs. 45%, and higher adherence was directionally associated with fewer failures, 14.06 vs. 19.46. *The practical issue is therefore not whether specifications should be explicit. They should. The issue is whether explicit obligations remain synchronized across specifications, prompts, tests, and code as requirements change.*

## IX. Implications for LLM-Assisted Development

Our findings shift the focus from better prompts to better specification continuity. Prior work has emphasized prompt engineering [139], [143], [108]. Our results suggest that prompts help when they preserve what the specification makes explicit. The challenge is to keep constraints, invariants, negative requirements, edge cases, tests, and code aligned as development proceeds. This has three implications. **First,** LLM-assisted tools should help developers identify missing constraints, invariants, assumptions, and negative requirements before code generation. **Second,** development environments should support specification-aware prompting and testing, so that written requirements become model interaction and validation. **Third,** requirement changes should trigger consistency checks across specifications, prompts, tests, and implementation, as partially propagated changes can become executable regressions.

Specification explicitness also matters for developer judgment. Participants without explicit obligations overestimated the correctness of generated code, whereas participants with explicit obligations produced better code but judged it more cautiously. This suggests that specifications not only guide the model but also provide developers with concrete checks to determine whether generated code should be trusted.

Future work should test these patterns in larger systems, with professional developers, longer maintenance tasks, and IDE-based or agentic LLM workflows. Tool support should help developers clarify underspecified behavior, generate tests from requirement elements, detect specification–code drift, and obligations missing from prompts/tests. This would move LLM assistance toward specification-aware development.

## X. Threats to Validity

As in any controlled experiment [144], our findings reflect the study design. Experimental control allowed us to isolate specification explicitness and support causal interpretation, but it limits the ecological breadth [145], [146].

**Internal validity.** Although participants were assigned using stratified randomization based on Python proficiency and LLM familiarity, residual differences in programming skill, requirements-engineering ability, or prior LLM experience may have affected outcomes. We mitigated this threat by using the same language, task family, process, timing, and evaluation protocol across groups. The treatment also changed both the artifact and the process: G2 and G3 participants externalized requirements before coding. Thus, our results should be interpreted as the effect of *explicit specification work*, including planning and reflection, rather than specification wording alone. Prompt-effort covariates and prompt–specification adherence analyses reduce, but cannot eliminate, this concern because they do not capture private reasoning or unrecorded decisions.

**Construct validity.** Hidden tests measure behavioral correctness and robustness, but not maintainability, readability, performance, security, or design quality. Failure modes rely on a predefined taxonomy, and some failures may span multiple categories. RQ2 codes if requirement elements are present, but not their precision or completeness. RQ4 measures calibration as the gap between predicted and observed hidden-test performance, which captures overconfidence but not broader developer understanding. Prompt-transcript labels also require interpretation; we mitigate this through independent coding and agreement reporting, but treat them as process indicators.

**External validity.** Participants were developers with prior requirements-engineering instruction and familiarity with `Python` and LLM-based development. This supports a controlled study of specification effects but limits direct generalization to industrial teams. Such experiments are common in empirical software engineering when limitations are acknowledged [147], [148], [149], [150]. The tasks were realistic but smaller than production systems, and the workflow used controlled ChatGPT-4o interaction rather than IDE-based assistants or autonomous agents. We therefore claim external relevance for the specification-explicitness mechanism, not a specific tool or setting.

**Conclusion validity.** The study detects large Phase I differences in correctness and failure structure, but weaker effects should be interpreted cautiously. This is especially true for RQ3, where only eight submissions regressed. We therefore emphasize confidence intervals, severity patterns, and practical magnitude rather than treating non-significant effects as evidence of no effect.

## XI. Conclusion

This paper shows that, in LLM-assisted development, specification explicitness is a control mechanism. It determines which behaviors are fixed before generation, which behaviors the model is left to infer, and which behaviors developers know to check. Our study shows that lightweight structure is not enough. In a controlled experiment with 72 participants, structured specifications only slightly reduced failures. Explicit behavioral obligations were different. They nearly halved hidden-test failures, reduced the number of distinct ways solutions failed, and most strongly reduced failures tied to implicit obligations. Under requirement change, explicitness did not remove regression, but it made regressions shallower. It also changed how developers judged generated code. Participants with explicit obligations produced better code but were less likely to overtrust it. The lesson is not to write longer specifications. The lesson is to make failure-relevant behavior explicit and keep it connected to prompts, tests, code, and change. For LLM-assisted development, specifications are not just input to the model. They shape generation, maintenance, and developer judgment.

## XII. Data Availability & Generative AI Statement

All anonymized data, analysis, and artifacts are available in an anonymous GitHub repository [122] and will be archived on Zenodo upon acceptance. Generative AI was part of the experimental methodology; outside the experiment, it was used only for language and grammar editing. The authors reviewed all revisions and take responsibility for the manuscript.